\documentclass{ws-ijmpa}
\usepackage{graphicx}
\usepackage{amsfonts}
\renewcommand{\d}{\partial}
\begin{document}
\def\draftnote{International Journal of Modern Physics A}
\title{ 
The influence of dark matter halo onto evolution of supermassive black hole}

\author{\footnotesize M. I. Zelnikov\footnote{ zelnikov@lpi.ru}}
\author{\footnotesize E. A. Vasiliev\footnote{eugvas@lpi.ru}}
\address{I.E.Tamm Theoretical Department\\
Lebedev Physical Institute \\
Leninsky pr.53, Moscow, Russia}

\maketitle

\begin{abstract}

The influence of dark matter (DM) on the growth of supermassive black holes (SMBHs)
is studied.
It is shown that gravitational scattering of DM particles on bulge stars leads 
to diffusion of DM in phase space \{$m, m_z, I$\} ($m$ denotes the angular momentum
and $I$ is the radial action).
Appropriate diffusion coefficients are calculated for different 
bulge models, and it is argued that the diffusion along
$m$ axis is the most important effect. 
It is shown that this process leads to noticeable flow of DM into
the black hole (BH), resulting in its power-law growth:
$M_{bh} \propto t^{9/16}$. Comparison with observational data shows that,
in principle, this effect may explain observed masses of SMBHs.
Special attention is paid to the corrections related to
the innermost region of BH gravitational influence 
and the diffusion along $I$ axis. Their influence on the BH growth law is 
shown to be negligible.

\end{abstract}

\section{Introduction}
 
The interaction of a supermassive black hole in a galaxy center with dark matter 
halo is already much investigated \cite{GS,MMH,UZK}. 
There an adiabatic invariant approach is applied for different halo structure.
The initial halo profiles are taken to be
self-similar (power-law) $\rho \sim r^{-\alpha} (0<\alpha<2)$, isothermal
($\rho \sim r^{-2}$) and NFW profiles\cite{NFW} ($\rho \sim \frac{\delta_c}{ r/r_s\,(1+r/r_s)^2}$).
In Ref.~\refcite{GS} the method of adiabatic invariant is used to calculate 
small changes in orbital parameters of particles caused by slow variation of 
gravitational potential due to the black hole growth.
This approach has two drawbacks:
absorption by BH is not taken into account, and appropriate change of DM
distribution function is neglected. The latter means that the loss cone
is always full, thus leading to overestimation of dark matter flow.

More correct approach is used in Refs.~\refcite{anton_sakh1,anton_sakh2}.
There the evolution of DM distribution function due to absorption by BH and
changes in BH mass and loss cone parameters are considered consistently. 
The authors come to the conclusion that for current values of black hole
masses the fraction of dark matter inside them is rather small.

This approach was developed in Ref.~\refcite{anton}, where the change of
dark matter distribution function outside the loss cone due to
diffusion in phase space was taken into account. It was demonstrated that this
diffusion caused by gravitational scattering of DM particles on stars
effectively refills the loss cone and determines the BH growth law.
This effect was shown to give reasonable estimate for observed black holes
masses.

The present work further develops the mentioned approach by accounting
for three other factors: change of gravitational potential in the vicinity of 
black hole, modification of star distribution in this region, and the 
conditions under which the diffusion becomes effectively two-dimensional.

The paper is organized as follows. In the first section we introduce our model 
of dark matter halo and the galactic bulge. In the second section we write down 
the kinetic equation which describes the evolution of dark matter halo in 
phase-space, and calculate several quantities related to star distribution and 
dark matter motion. Calculation of the diffusion coefficients is performed in 
the third section. Also estimated are the diffusion timescales and corrections 
related to the central region. In the fourth section the solution of the 
diffusion equation in one-dimensional form is presented, the black hole growth 
law is established, and the corrections to the DM flow due to two-dimensional 
diffusion and the central region are investigated. It is shown that these 
corrections in most cases play no significant role in the diffusion process.
Finally, in the section 5 we present some numerical results and discuss them.

\subsection{Dark matter in galaxies}

While the most part of matter in the Universe is now proved to be dark, the
nature of the dark matter still remains unclear. In most models, however, the
significant or dominant part of dark matter is taken to consist of cold dark
matter (CDM), i.e. non-relativistic particles interacting only gravitationally.
This leads to the absence of thermodynamical equilibrium in DM gas and 
the necessity of kinetic approach.
Following this approach it was shown that the initial inhomogenities 
in DM distribution grow and form spherically-symmetric structures -- 
non-dissipative gravitational singularities 
(NGSs) of different scales, having the same internal structure. General analytic 
theory of NGS formation \cite{GZ} (consistent with numerical simulations \cite{FM,Moore}
and other analytic theories \cite{WBD}) 
predicts that all these structures have similar density profiles described by the 
following formula:
\begin{equation}  \label{rho_init}
\rho(r) = A\,r^{-\xi} \;,\; \xi=12/7.
\end{equation}
The density profiles of DM haloes appearing in numerical simulations (see 
Refs.~\refcite{NFW,Moore}) have different forms, typically shallower in the centre. 
While being more realistic in the outer parts than the profile (\ref{rho_init})
due to hierarchical formation mechanism, they cannot describe well the very 
inner region because of limited spatial resolution,
thereby allowing different values of the inner slope \cite{Klypin1,Navarro2003}
ranging from $-1$ to $-1.5$, all of which are within the errors
in the simulations. Furthermore, observations do not exclude cuspy density 
profiles \cite{Klypin2}, but still are unable to determine the inner slope 
precisely. Therefore, the adopted profile does not contradict observations
and is convenient for a number of reasons (simplicity of distribution function
and tight relation between angular momentum and energy). 

As initial inhomogenities are not spherically symmetric, the particles of 
NGS possess angular momenta (though total angular momentum of NGS is zero).
High degree of contraction at the nonlinear stage makes these momenta 
relatively small compared to maximal possible angular momenta for corresponding 
orbital sizes. It turns out that angular momentum of a particle is related to its 
radial action $I$ by the formula: $m^2 = {l_0}^2 {I}^2$,
where $I = \frac{1}{\pi} \int_{r_-}^{r_+} v_r(r)\,dr$,
$v_r=\sqrt{2(E-\Psi(r))-\frac{m^2}{r^2}}$ is the radial velocity, $\Psi(r)$ is 
the gravitational potential, $r_\pm$ are the turnpoints,
and  $l_0 \simeq 0.1$ is a small parameter. 


In this case, the distribution function of DM in canonical variables
$I$, $m$, $m_z$ has the following form:
\begin{equation}  \label {f_init}
f(I, m, m_z) = f_0\,{I}^{1/8}\,\delta(m^2-{l_0}^2 {I}^2)
\end{equation}
The average value of $m$ is zero, though the average $m^2$ is nonzero.

The formation of NGS of galactic scale is followed by the formation of galaxy 
itself due to infall of baryonic matter into gravitational well of NGS, its 
subsequent cooling and formation of galactic structures -- disc, bulge and halo.
After this stage the total gravitational potential is no longer defined by DM; 
conversely, in the central region it is mainly directed by baryons. 
Galaxy formation process is slow enough as compared to the dynamical time of DM 
particles, so that the gravitational potential evolution is adiabatic.
This fact ensures that radial action is integral of motion (adiabatic invariant).
It is defined by equation
\begin{equation}  \label{I_r}
I(E,m) = \frac{1}{\pi}\int_{r_-}^{r_+} dr \sqrt{2(E-\Psi(r))-\frac{m^2}{r^2}}
\end{equation}
where $r_-$, $r_+$ are minimal and maximal distances of a DM particle from
center; $r_- \ll r_+$ due to small angular momentum ($m \ll I$).

Our main task is to study the central region of a galaxy -- the bulge, which 
we may consider to be spherically symmetric. Under this assumption angular momentum 
is also conserved, and we obtain that the DM distribution function written in
variables $I$, $m$, $m_z$ does not depend on time.
Spatial distribution of DM is given by relation
\begin{equation}
\rho(r) = (2\pi)^3 \frac{1}{4\pi r^2}\int_0^\infty dm \int_{-m}^{+m} dm_z 
\int_\Omega dE \frac{\sqrt{2}}{\pi}\frac{1}{2\sqrt{E-\Psi-\frac{m^2}{2r^2}}}
f(E,m) \;,
\end{equation}
$\Omega$ is the energy interval where 
the expression under the radical sign is
non-negative.

Let us assume that the density and the potential of the bulge have
power-law profiles:
$$
n(r) \propto r^{-(2-\alpha)} \;,\quad M_b(r) \propto r^{\alpha+1} \;,\quad 
\Psi(r) \propto r^\alpha \;,
$$
Then one can show \cite{anton_sakh1,GS} that the radial action may be approximated
(with the accuracy better than 8\%) in a factorised form
\begin{eqnarray}  \label{I_fact}
&&I = I_0(r_+)\, C(\mu)\;, \quad \mu=\frac{m}{I_0} \\
&&I_0 = \sqrt{G\,M_b(r_+)\,r_+} \nonumber \\
&&C(\mu) = \frac{\sqrt{2}}{\pi} \int^1_\frac{r_-}{r_+} d\chi \left[ 
  \int_\chi^1 \xi^{\alpha-1} d\xi - \frac{\mu^2}{2} 
  \left(\frac{1}{\chi^2}-1\right) \right]^{1/2} \nonumber
\end{eqnarray}
where function $C(\mu)$ at small argument $\mu$ is linear:
\begin{equation}  \label{C_mu}
C(\mu) \approx C(0)-b_\alpha\mu \;,\quad C(0)>0\;,\;\;b_\alpha \sim 1
\end{equation}

Hence the DM density profile is also power-law:
\begin{equation}
\rho(r) = A' r^{-\xi'} \;,\quad \xi' = -\frac{15}{8}+\frac{9}{16}\alpha
\end{equation}

Let us now overview the bulge structure and proceed to estimation of 
DM fraction in the bulge.

\subsection{Bulge overview}

Bulge is the central part of a galaxy,
for our Galaxy its radius being about 1 kpc and mass about $10^{10} M_\odot$ \cite{bulge}. 
We assume it to be spherically symmetric.
Bulge density profile can be derived from the dependency of stars velocity 
dispersion on distance to galactic center; we take it to be power-law:
$n(r) = \tilde \eta_b r^{-(2-\alpha)}$. For isothermal bulge, i.e. having
uniform velocity dispersion, $\alpha=0$. Milky Way bulge may be approximated
with velocity dispersion $\sigma \propto r^{1/4}$ for $r \le 50\mbox{ pc}$
\cite{Tremaine}, which corresponds $\alpha=0.5$.

It should be noticed that currently available observations of distant galaxies
usually do not have enough resolution to derive velocity dispersion on scales 
less than a hundred parsecs \cite{Tremaine}. We assume their 
bulges to be isothermal in central parts, but one can show that the exact form 
of density profile does not matter much; only its integral characteristics 
affect the final result.

Generally, in the center of a galaxy a compact object is located, which is 
assumed to be a supermassive black hole (for a recent review see Ref.~\refcite{smbh}).
For our own Galaxy it is now proved that the compact object Sgr A* is a black 
hole with mass $M_{bh} \simeq 2.9\cdot 10^6 M_\odot$ \cite{genzel}. 
The innermost star cluster has the following density profile: 
\begin{equation}  \label{center}
n(r) = n_0 \left(\frac{r}{r_0}\right)^{-3/2} \mbox{, where } 
n_0=10^8 \frac{1}{\mbox{pc}^3}\;,\; r_0=0.02\mbox{ pc \cite{genzel2}.}
\end{equation}
Hence the mass of this cluster is $M(r) = \int_0^r M_s n(r')\,4\pi 
r'^2\,dr' = \frac{8\pi}{3}M_s n_0{r_0}^3 \left(\frac{r}{r_0}\right)^{3/2}$. 
For the sake of simplicity we have taken the masses of all stars to be equal
$M_s=M_\odot$.
Let us define the influence radius $R_0$ of BH by the condition that the total mass 
of stars inside this radius equals that of the BH:
\begin{equation}  \label{R_0_now}
R_0 = \left( \frac{M_{bh}}{M_s} \frac{3}{8\pi\,n_0{r_0}^{3/2}}\right)^{2/3} 
\approx \mbox{1 pc at the moment.}
\end{equation}

So we may assume that inside this radius the gravitational potential is 
dominated by BH, while outside it the influence of BH is negligible.

Notice that BH and bulge evolution changes values of $n_0$ and $R_0$.
If we assume that the star density is continuous at $R_0$ while $R_0$ increases
due to BH growth, then the following relation 
between $M_{bh}$ and $R_0$ arises:
\begin{equation}  \label{R_0}
M_{bh} = \frac{8\pi}{3} M_s\, \tilde \eta_b\,{R_0}^{1+\alpha}.
\end{equation}
Here $\tilde \eta_b$ is the coefficient in the bulge density profile; for
isothermal bulge with velocity dispersion $\sigma_0$
$\tilde \eta_b = \frac {{\sigma_0}^2}{2\pi\,G\,M_s}$ (see (\ref{n_iso})).

\subsection{Dark matter in the bulge}

The total mass of Milky Way's dark matter halo can be assessed to be $M_H \sim 
10^{12}M_\odot$, its radius to be $R_H \sim$ 100 kpc \cite{halo}. 
From (\ref{I_fact}) we derive adiabatic invariant for particles with 
$R_+=R_H$ to be $I_{max}=0.35 (G\,M_H\,R_H)^{1/2}$.
If we suppose that the initial distribution (\ref{f_init}) can be extrapolated 
up to $R_+=R_H$, then tha total halo mass is
\begin{equation}  \label{M_I}
M_H=\frac 8 9 (2\pi)^3 f_0 {I_{max}}^{9/8}\;,
\end{equation}
and hence
\begin{equation}  \label{f_0}
f_0 = 3.5 \frac{1}{(2\pi)^3} \frac{{M_H}^{7/16}}{(G\,R_H)^{9/16}}\;.
\end{equation}
Then the value of $f_0$ for our Galaxy is $6\cdot 10^8 \mbox{g} 
({\textstyle \frac{\mbox{cm}^2}{\mbox{s}}})^{-9/8}$.

In this case the relative amount of DM inside the bulge is given by
\begin{equation}  \label{Ups}
\Upsilon = \frac{M_{DM}}{M_B} = \frac{M_H}{(G\,M_H\,R_H)^{9/8}} \frac{(G\,M_B\,R_B)^{9/8}}{M_B}
= \left(\frac{M_H}{M_B}\right)^{\frac{7}{16}} \left(\frac{R_B}{R_H}\right)^{\frac{9}{16}}
\sim 1.
\end{equation}
This means that for the model distribution (\ref{f_init}) the mass of dark
matter inside bulge is comparable to the baryonic mass. The same relation is
valid in the innermost region (BH domain of influence).
However, the distribution (\ref{f_init}) is applicable only in some
vicinity of the galactic center, and hence the actual values of $f_0$ and
$\Upsilon$ can deviate from the assessments (\ref{f_0}, \ref{Ups}).

Observations and modelling of mass distribution in the Galaxy do not give tight 
restrictions on the amount of DM in the bulge. Depending on the model, $\Upsilon$
may comprise from $1/4$ to $1/3$ of the total mass inside the central 3 kpc
region \cite{Klypin2}, but not the major fraction of the bulge mass. 
So, for the final calculations of the BH mass we should decrease the value of
$f_0$ (\ref{f_0}) by a factor of 3--4 or more.

\section{Problem definition and particle motion parameters}

Our goal is to determine the possible fraction of DM in BH, and to derive the 
growth law of BH due to absorption of DM. 
Firstly, one can prove that the direct capture of particles with momenta less than
$m_g = \frac{4\,G\,M_{bh}}{c}$ does not lead to significant growth of BH since 
their mass calculated from distribution (\ref{f_init}) with current value of 
$m_g$ is several orders of magnitude less than $M_{bh}$ \cite{anton_sakh2}. 

The simplest process which changes the distribution function of DM is 
gravitational scattering of DM particles on stars identical to 
Coulomb scattering in plasma.

\subsection{Kinetic equation}

The dark matter distribution function satisfies the following equation
\begin{equation}  \label{kin_eq_spatial}
\frac {\d f (\vec r, \vec v, t)}{\d t} + \{H_0, f\} = St\{f\} \;,
\end{equation}
where $H_0$ is the Hamilton function for gravitational interaction,
$\{,\}$ is the Poisson bracket,
$St\{f\}$ is the collision term in Landau form \cite{Landau}:
\begin{eqnarray}
&&St\{f\}=\frac{\d}{\d(\mu v_i)} \int\left[ f(\vec v) \frac{\d f'(\vec v')}{\d (\mu v'_j)}
  - f'(\vec v') \frac{\d f(\vec v)}{\d (\mu v_j)}\right] B_{ij} \,d^3 v' \;,\\
&&B_{ij} = \frac{2\pi G^2 {\mu'}^2 \mu^2 L_c}{|u|} 
  \left(\delta_{ij}- \frac{u_i u_j}{u^2}\right) \;.\nonumber
\end{eqnarray}

Here entities with primes refer to stars and other to DM particles; masses are 
denoted as $\mu, \mu'$ to avoid confusion with momentum.
$\vec u=\vec v-\vec v'$ is the relative velocity, $L_c \sim 15$ is the
Coulomb logarithm.
Since the mass of a star $M_s \equiv \mu' \gg \mu$, we neglect the first term in equation:
\begin{eqnarray}  \label{Stvr}
&&St\{f\}=\frac{\d}{\d v_i}\left[ W_{ij} \frac{\d f}{\d v_j}\right] \;,\\
&&W_{ij} = 2\pi G^2 {M_s}^2 L_c \int \frac{u^2 \delta_{ij} - u_i u_j}{u^3} 
f'(\vec v', r) d^3 v'  \;.\nonumber
\end{eqnarray}

As soon as collision frequency of DM particles with stars is much less than their 
orbital motion frequency, one can rewrite the equation (\ref{kin_eq_spatial})
in the form averaged over a period in action-angle variables
\{$I_k, \phi_k$\}; $\{I_k\} = \{I, m, m_z\}$:
\begin{equation}  \label{kin_eq}
\frac {\d f (\{I_k\}, t)}{\d t} = St\{f\}\;.
\end{equation}
Then the equation (\ref{Stvr}) for collision term looks as follows:
\begin{eqnarray}  \label{St3}
&&St\{f\}=\frac{\d}{\d I_k}\left[ R_{kl} \frac{\d f}{\d I_l}\right] \;,\\
&&R_{kl} =\frac{1}{(2\pi)^3} \int d^3\phi\, \frac{\d I_k}{\d v_i} 
\frac{\d I_l}{\d v_j}\, W_{ij}  \;.\label{Rkl3}
\end{eqnarray}

The star distribution around a SMBH and their diffusion towards the disruption
boundary was studied in Refs.~\refcite{BW,LS}. Two-dimensional Fokker-Planck equation was 
written in coordinates $E$ and $m$, and solved analytically with certain 
simplifications, introducing the notion of energy-dependent loss-cone.
A fraction of stars with small angular momenta is eliminated during one 
dynamical time, and the diffusion along energy axis leads to capture of 
stars with critical (minimal) energy. We use radial action $I$ instead of 
energy, because diffusion along $I$ axis does not lead to capture; therefore
the absorption process looks more clear. Also, our distribution function is
time-dependent and the black hole mass changes in time.

Observations show that the distribution function (DF) of stars
may be considered as isotropic, i.e. not depending on momentum, even in
the central region \cite{genzel}.
For isotropic DF the tensor $W_{ij}$ takes the following form \cite{anton}:
\begin{eqnarray}
W_{ij}&=&A(E,r)\delta_{ij} - B(E,r)\frac{v_i v_j}{v^2} \;, \\
A & = & \frac{16\pi^2}{3} G^2 {M_s}^2 L_c \int\limits_{\Psi(r)}^\infty 
  dE' f'(E') \left\{ \begin{array}{lcl} 1&,&E \le E' \\ 
  \frac{3}{2}\frac{v'}{v}(1-\frac{v'^2}{3\,v^2}) &,&E>E' \end{array}\right. \\
A-B &=&  \frac{16\pi^2}{3} G^2 {M_s}^2 L_c \int\limits_{\Psi(r)}^\infty 
  dE' f'(E') \left\{ \begin{array}{lcl} 1&,&E \le E' \\ 
  (\frac{v'}{v})^3 &,&E>E' \;\;.\end{array}\right. 
\end{eqnarray}

From (\ref{I_fact}, \ref{C_mu}) it follows that adiabatic invariant may be represented as 
follows:
\begin{equation}  \label{IJm}
I(E,m) \approx J(E) - b_\alpha m\;.
\end{equation}
Note that in the case of Coulomb ($\alpha=-1$) and oscillator ($\alpha=2$) potential 
the constant $b_\alpha=1$, and in the case of isothermal potential
(or close to it) $b_\alpha \approx 0.6$. 

It is convenient to perform linear variable change: $\{I, m, m_z\} \to \{J, m, m_z\}$
Since it is linear, the expression (\ref{Rkl3}) for tensor $R_{kl}$ does not change.
In addition, due to small value of parameter $l_0$, the initial distribution function 
(\ref{f_init}) has the same form in new variables: 
\begin{equation}  \label{f_init_Jm}
f_i(J,m) = f_0 J^{1/8} \delta(m^2-{l_0}^2 J^2)
\end{equation}

The initial distribution function does not depend on $m_z$, hence the solution of
kinetic equation (\ref{kin_eq}) will not depend on $m_z$ neither.
We rewrite the expression (\ref{St3}) for collision term as follows:
\begin{equation}  \label{Stf}
St\{f\} = \frac 1 m \frac \d{\d m} m 
  \left( R_{22}\frac{\d f}{\d m} + R_{12}\frac{\d f}{\d J}\right) + \frac \d{\d J}
  \left( R_{12}\frac{\d f}{\d m} + R_{11}\frac{\d f}{\d J}\right) \;,
\end{equation}
where the diffusion coefficients (\ref{Rkl3}) are the following \cite{anton}:
\begin{eqnarray}  \label{Rkl}
R_{11} &=& \left(\frac{\d J}{\d E}\right)^2 \left< (A-B)v^2 \right> \;,\\
R_{12} &=& \left(\frac{\d J}{\d E}\right)   \left< (A-B) m  \right> \;,\nonumber\\
R_{22} &=& \left< A\,r^2 - B\frac{m^2}{v^2} \right> \;.\nonumber
\end{eqnarray}
Here the averaging over angle variables is defined as
\begin{eqnarray}
&&\langle X\rangle = \frac 2 T\int_{r_-}^{r_+} \frac{dr}{v_r}\,X \;, \\
&&T=2\int_{r_-}^{r_+}\frac{dr}{v_r}\;,\quad 
v_r=\sqrt{2(E-\Psi(r))-\frac{m^2}{r^2}} \;,
\end{eqnarray}
where $T$ is the particle oscillation period, $v_r$ is its radial velocity.

Before proceeding to the solution of the kinetic equation (\ref{kin_eq}, \ref{Stf})
one should define the distribution function of stars.

\subsection{Distribution of stars}

We assume that the density profile of stars in the bulge is power-law:
$n(r) = n_0 \left(\frac{r}{r_0}\right)^{-\gamma}$, and the gravitational potential
is also power-law $\Psi(r)=\Psi_0 r^\alpha$. In this case isotropic distribution
function of stars can be written as a power-law dependence on energy \cite{PF}:
\begin{equation}  \label{fprime}
f'(v',r)=F_0 E^{-\beta}\;,\quad E=\frac{v'^2}2+\Psi(r) \;,
\end{equation}
$$
n(r) = \int_0 f'(v',r)\,d^3v' = 
\int_0^\infty F_0 (\frac{v'^2}2 + \Psi_0 r^\alpha)^{-\beta}\, 4\pi v'^2\,dv' =
$$
\begin{equation}  \label{alphagamma}
= \frac{\Gamma(\beta-\frac 3 2)}{\Gamma(\beta)} (2\pi)^{3/2} F_0 \,
{\Psi_0}^{\frac 3 2 - \beta}\, r^{\alpha\,(\frac 3 2 - \beta)}\;
\mbox{, hence }
\gamma=(\beta-\frac 3 2) \alpha.\qquad\qquad\qquad
\end{equation}

We consider two particular cases: a) self-consistent potential of stars in the bulge
(from observations \cite{faber} it follows that the star distribution in the 
bulge is close to isothermal, so we take the power-law index $\gamma$ close to 2)
and b) the central region of bulge where potential is dominated by the black hole.

In the first case we have
$$
\frac{d\Psi(r)}{dr} = \frac{4\pi\, G\,M(r)}{r^2} = \frac{4\pi}{3-\gamma} G\, 
M_s\,n_0\,r_0^\gamma r^{1-\gamma}  \;\mbox{, hence}
$$
\begin{equation}  \label{Psi0conf}
\Psi=\Psi_0 r^\alpha \;,\qquad 
\Psi_0 = \frac{4\pi\,G\,M_s\,n_0\,r_0^\gamma}{(3-\gamma)(2-\gamma)} \;,\quad 
\gamma=2-\alpha \;,
\end{equation}
\begin{equation}  \label{sigmaconf}
\sigma=\sigma_0 r^{\alpha/2} \mbox{ is the velocity dispersion, }
{\sigma_0}^2 = \Psi_0 {\textstyle \frac{\alpha}{2(1-\alpha)}} \;,
\end{equation}
\begin{equation}  \label{F0conf}
f'(E)=F_0E^{-\beta} \;,\qquad 
F_0 = \frac{\alpha(\alpha+1)}{4\pi\,G\,M_s(2\pi)^{3/2}}
\frac{\Gamma(\beta)}{\Gamma(\beta-\frac 3 2)} \,{\Psi_0}^{\frac 2\alpha} \;,\qquad 
{\textstyle \beta=\frac 2 \alpha + \frac 1 2}\;.
\end{equation}

In the limit $\alpha \to 0$ these expressions describe isothermal star cluster.
Let us rewrite them in this particular case separately:
\begin{eqnarray}  \label{Psiiso}
\Psi(r) &=& \Psi_0 \ln\frac{r}{r_0}\;,\quad \Psi_0=2{\sigma_0}^2 
  \;,\quad \sigma_0 \mbox{ is the velocity dispersion,} \\
n(r)    &=& n_0 \left(\frac{r}{r_0}\right)^{-2} \;,\qquad \label{n_iso}
  n_0{r_0}^2 = \frac{{\sigma_0}^2}{2\pi G M_s} \;,\\
f'(E)   &=& F_0 \exp(-\frac{E}{{\sigma_0}^2}) \;,\qquad 
  F_0=n_0 (2\pi{\sigma_0}^2)^{-3/2} \;.
\end{eqnarray}

In the second case the particle dynamics is governed by the Coulomb potential
of the black hole.
\begin{equation}  \label{Psicoul}
\Psi(r)=-\frac{G\,M_{bh}}{r} \;;
\end{equation}
Restricting our consideration to the case $\gamma=3/2$ which follows from observations \cite{genzel2}, 
we obtain from (\ref{alphagamma}) $\beta=0$. Hence the distribution function does not
depend on energy and equals $F_0$ at $E<0$:
\begin{equation}  \label{F0coul}
f'(E)=F_0=\frac{3}{8\sqrt{2}\pi(G\,M_{bh})^{3/2}}\frac{3\,M_{bh}}{8\pi\,M_s\,{R_0}^{3/2}}
\;,
\end{equation}
where $M_{bh}$ and $R_0$ are related by (\ref{R_0}).

\subsection{Dark matter motion parameters}

In this section we present the motion parameters for DM particles for
practically important cases.
\begin{eqnarray*}
I(E,m) &=& \frac 1 \pi \int_{r_-}^{r_+} \sqrt{2(E-\Psi(r))-\frac{m^2}{r^2}}\, dr
\mbox{ -- adiabatic invariant; } \\
I(E,0) &\equiv& J(E) \;,\;\; r_\pm \mbox{ -- turnpoints;} \\
T(E,m) &=& 2\pi \frac{\d I}{\d E}  \mbox{ -- particle oscillation period;} \\
\frac{r_-}{r_+} &=& \chi_{min} \ll 1 \;, \quad \frac{m}{J} = \mu \ll 1\;.
\end{eqnarray*}

For power-law potential (\ref{Psi0conf}):
\begin{eqnarray}
J(E) &=& C_\alpha\, \frac{\sigma_0\, {r_+}^{1+\frac{\alpha}2}}{\sqrt{\pi}} \;, \quad
  C_\alpha = \textstyle{\frac{1}{\sqrt{\pi}}} \int\limits_0^1 d\chi
  \sqrt{{\textstyle\frac{4(1-\alpha)}{\alpha}} (1-\chi^\alpha)} 
  \,\approx 1 \scriptstyle{- 0.8\alpha} ,\\
T(E) &=& C_\alpha\, \frac{ 1 + \frac \alpha 2}{1-\alpha} \, 
  \frac {\sqrt{\pi} \, {r_+}^{1-\frac \alpha 2}}{\sigma_0} \;,\\
r_+ &=& \left( \frac{E}{2{\sigma_0}^2} \frac{\alpha}{1-\alpha} \right)^{1/\alpha}
  \;,\qquad  \chi_{min} = C_\alpha \sqrt{\frac{\alpha}{4\pi(1-\alpha)}}\,\mu  \;.\label{chiminconf}
\end{eqnarray}

In the limit $\alpha \to 0$ we obtain formulas for isothermal potential (\ref{Psiiso}):
\begin{eqnarray}
J(E) &=& \frac{1}{\sqrt{\pi}} \sigma_0 r_+ \;,\\
T(E) &=& \sqrt{\pi} r_+ / \sigma_0 \;,\\
r_+  &=& r_0\,\exp\left(\frac{E}{2{\sigma_0}^2}\right)  \;,\qquad
  \chi_{min} = \frac{0.25\mu}{\sqrt{1.3-\ln\mu}}\;.
\end{eqnarray} 

For Coulomb potential (\ref{Psicoul}):
\begin{eqnarray}
J(E) &=& \frac{1}{\sqrt{2}} \sqrt{G\,M_{bh}\,r_+} \label{Jcoul}\;,\\
T(E) &=& \frac{\pi}{\sqrt{2}} \frac{G\,M_{bh}}{|E|^{3/2}} \;,\\
r_+  &=& \frac{G\,M_{bh}}{|E|} \;,\qquad \chi_{min} = \frac{\mu^2}{2}
\label{chimincoul}\;.
\end{eqnarray}

\section{The diffusion coefficients}

Two opposite cases should be distinguished: for particles with apocenter
distances $r_+(J) < R_0$ ($R_0$ is the radius of BH's domain of influence (\ref{R_0}))
we calculate coefficients using Coulomb potential, and for other 
particles we use the expressions for self-consistent bulge potential (this 
corresponds to assumption that these particles spend the most part of orbital period 
outside BH's domain of influence).

\subsection{The diffusion coefficients for bulge}  \label{coef_bulge}

Below we present the quantities $A, A-B$ for power-law (including isothermal) star distribution:

\newcommand{\AB}{\Omega}
\begin{eqnarray}  \label{fncA}
A  &=& \frac{4}{3\sqrt{2\pi}} G\,M_s\,L_c\,\sigma(r)\,r^{-2} \cdot 
\widetilde A_\alpha(\tilde v) \;,\quad  \tilde v = \frac{v(r)}{\sigma(r)} \;,\\
A-B&=& \frac{4}{3\sqrt{2\pi}} G\,M_s\,L_c\,\sigma(r)\,r^{-2} \cdot 
\widetilde \AB_\alpha(\tilde v)  \;,\label{fncAB}
\end{eqnarray}
where
\begin{eqnarray*}  
\widetilde A_\alpha(\tilde v) &=& \left[ 
 \frac{{(\tilde v^2}/{2\varsigma}+1)^{-(\beta-1)}}{\scriptstyle\beta-1} +
 \frac{\tilde v^2}{2\varsigma} F\left(\frac 3 2, \beta; \frac 5 2; -\frac{\tilde v^2}{2\varsigma}\right) -
 \frac{\tilde v^2}{10\varsigma}F\left(\frac 5 2, \beta; \frac 7 2; -\frac{\tilde v^2}{2\varsigma}\right) \right]
\times \\
&\times& \frac{\alpha(\alpha+1)\,\sqrt{\varsigma}}{2} 
\frac{\Gamma(\beta)}{\Gamma(\beta-\frac 3 2)}  \;,\\
\widetilde \AB_\alpha(\tilde v) &=& \left[ 
 \frac{({\tilde v^2}/{2\varsigma}+1)^{-(\beta-1)}}{\scriptstyle\beta-1} +
 \frac{\tilde v^2}{5\varsigma} F\left(\frac 5 2, \beta; \frac 7 2; -\frac{\tilde v^2}{2\varsigma}\right) \right]
\times \frac{\alpha(\alpha+1)\,\sqrt{\varsigma}}{2} 
\frac{\Gamma(\beta)}{\Gamma(\beta-\frac 3 2)} \;,
\end{eqnarray*}
$$
\varsigma = 2\frac{1-\alpha}{\alpha}=\frac{\Psi_0}{{\sigma_0}^2} \;, \qquad
F(..,..;\dots;.)\mbox{ is the hypergeometric function,}
$$
and for isothermal profile
\begin{eqnarray*}
\widetilde A_0(\tilde v) &=& \frac 3 2 \left[
\frac{\exp(-\frac{\tilde v^2}{2})}{\tilde v^2} + \frac{\tilde v^2-1}{\tilde v^3}
\Phi(\tilde v) \right] \;, \\
\widetilde \AB_0(\tilde v) &=& 3 \left[
-\frac{\exp(-\frac{\tilde v^2}{2})}{\tilde v^2} + \frac{1}{\tilde v^3}
\Phi(\tilde v) \right] \;,\qquad
\begin{array}{l}
\Phi(x) = \int_0^x \exp(-\frac{t^2}{2}) dt \\
\mbox{is the probability integral.} \end{array}
\end{eqnarray*}

\begin{figure}
$$\includegraphics{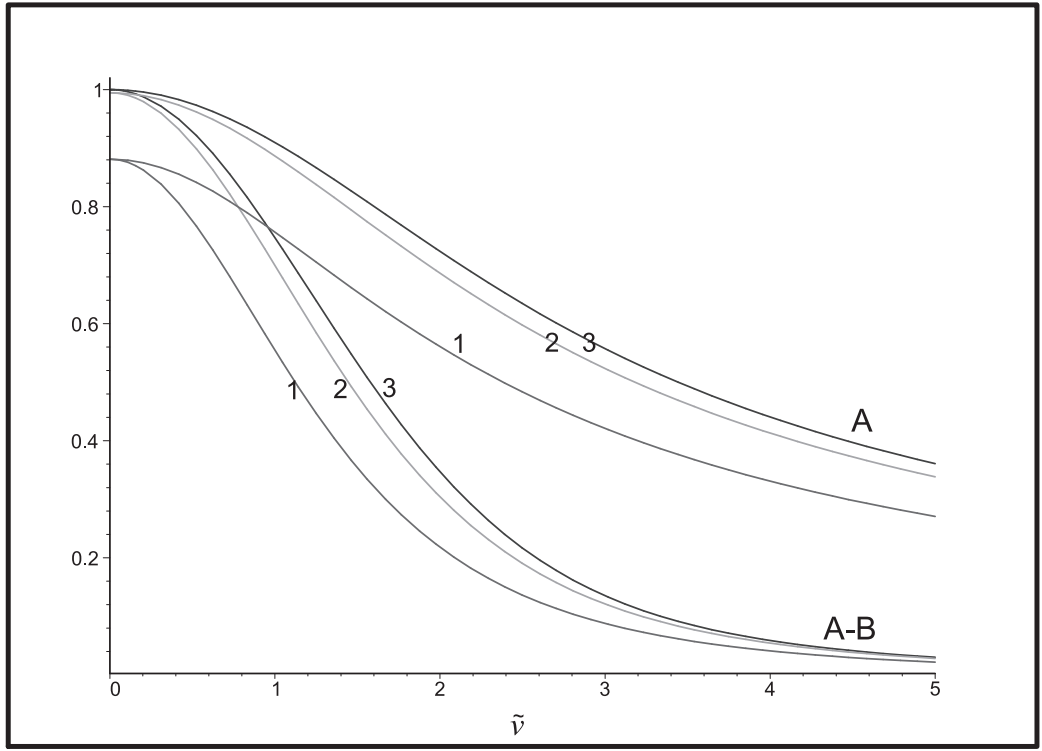} $$
\caption{ 
The graphs of functions $\widetilde A(\tilde v)$ (coefficient $A$, (\ref{fncA})), 
$\widetilde \AB(\tilde v)$ (coefficient $A-B$, (\ref{fncA})) 
for different potentials:
1) power-law ($\Psi \propto r^{\alpha}$), $\alpha=1/2$; 
2) power-law, $\alpha=1/4$; 3) isothermal ($\alpha=0, \Psi \propto \ln r$).
}  \label{A_B}
\end{figure}

The graphs of functions $\widetilde A_\alpha(x)$, $\widetilde \AB_\alpha(x)$ 
are displayed on fig.\ref{A_B}

Now let us calculate the diffusion coefficients $R_{11}$, $R_{12}$, $R_{22}$
according to (\ref{Rkl}).

First of all, we are interested in low values of momentum, so we neglect the 
second term in $R_{22}$, which is $\sim \mu^2$ times smaller than the first one.
\begin{equation}  \label{R22b_int}
R_{22} = \frac{8}{3\sqrt{2\pi}} G\,M_s\,L_c\,r_+\,T^{-1}
  \int_{\chi_{min}}^1 \frac{d\chi}{\tilde v} \widetilde A(\tilde v) \;, \qquad
\chi = \frac{r}{r_+} \,
\end{equation}
$$
\tilde v = \left\{ \begin{array}{lcl} 
 \sqrt{\frac{4(1-\alpha)}{\alpha}} \sqrt{\chi^{-\alpha}-1} &,& \alpha>0 \vspace{2pt}\\
 \sqrt{-4 \ln(\chi)} &,& \alpha=0  \;\;.\end{array} \right.
$$
We take the lower limit of integration in $\chi_{min}$ to be zero, since at low $\chi$ 
(high $\tilde v$) the integrand is small. Then we have for $R_{22}$ the following expression:
\begin{equation}  \label{R22b}
R_{22} \approx 0.46\, G\,M_s\,L_c\,{\sigma_0}^\frac{1}{1+\alpha/2}\,J^\frac{\alpha}{2+\alpha}
\;.
\end{equation}
This conforms the value obtained in Ref.~\refcite{anton}. Note that 
$R_{22}$ does not depend on $m$ and weakly depends on $J$, not depending on $J$ 
at all in isothermal case.

For $R_{11}$ we have
\begin{equation}  \label{R11b_int}
R_{11} = \frac{8}{3\sqrt{2\pi}} G\,M_s\,L_c\,r_+ T^{-1}
  \left(\frac{T \,\sigma_0\, {r_+}^{\alpha/2-1}}{2\pi}\right)^2 
  \int_{\chi_{min}}^1 d\chi\,\chi^{-2+\alpha}\, \tilde v\, \widetilde  \AB(\tilde v) 
  \;.
\end{equation}
The integral diverges at lower limit, so we take the value (\ref{chiminconf})
for $\chi_{min}$.
For the isothermal case ($\alpha=0$) it is possible to obtain an asymptotic 
approximation of this integral at low $\chi_{min}$. 
Since $\widetilde \AB_0(\tilde v) \approx 
\frac{3\sqrt{\pi}}{\sqrt{2}\, \tilde v^3}$, the integral approximately equals
$(-\chi_{min} \ln\chi_{min})^{-1} \sim 1/\mu$. A similar consideration for
$\alpha>0$ leads to an estimate of the integral as $\mu^{-1+2\alpha}$. 
The expression in brackets before the integral weakly depends on $\alpha$ 
and equals $1/{2\sqrt{\pi}}$ for $\alpha=0$. Then
\begin{equation}  \label{R11b}
R_{11} \approx R_{22} \cdot 0.1 \left(\frac J m \right)^{1 - 2\alpha} \;.
\end{equation}

The last coefficient $R_{12}$ can be represented as $R_{22}\cdot K_\alpha(\mu)$, 
where $K_\alpha(\mu) \to 0$ at $\mu \to 0$. So one can neglect the term with 
$R_{12}$ in (\ref{Stf}) at low $\mu$.

\subsection{The diffusion coefficients for central region}

We restrict the calculation to the case when the density of stars
$n(r) \propto r^{-3/2}$. Then
$$
R_{22} = \frac{32\pi^2}{3\sqrt{2}} G^2 {M_s}^2 L_c F_0 {r_+}^3 |E|^{1/2} 
  T^{-1}  \int_{\chi_{min}}^1 \frac{\chi^2\,d\chi}{\sqrt{\frac 1 \chi -1}}
\left( \frac{4}{5\chi} - \frac 1 5 \right) \;.
$$
The integral weakly depends on $\chi_{min}$ at low $\chi_{min}$ and equals 
$\approx 1.14$. Finally, we have
\begin{equation}  \label{R22c}
R_{22} = 2.4 G^{1/2}\,M_s\,{M_{bh}}^{-1/2}\,{R_0}^{-3/2}\, L_c\, J^2 \;.
\end{equation}
Comparing (\ref{R22c}) with expression (\ref{R22b}) for bulge, we conclude 
that they coincide at $J=J(R_0)$ (i.e. at the boundary of central region of 
the black hole influence).
This means that the expressions for two limiting cases are consistent with each other.

\begin{equation}  \label{R11c_int}
R_{11} = \frac{16}{3\sqrt{2}} G^2 {M_s}^2 L_c F_0 r_+ |E|^{3/2} T
\int_{\chi_{min}}^1 d\chi \sqrt{\textstyle\frac{1}{\chi}-1}\,(\frac{2}{5\chi}+\frac{3}{5})
\end{equation}
Here the integral approximately equals $0.8/\sqrt{\chi_{min}}$; 
so with the aid of
(\ref{chimincoul}) we obtain
\begin{equation}  \label{R11c}
R_{11} = R_{22} \cdot 0.25\,\frac{J}{m}
\end{equation}
Similarly to the previous case we get that $R_{12} \simeq {\rm const}\, 
R_{22}\, \frac{m}{J}$, so it can be neglected at low $m$.

\subsection{Diffusion timescale estimate}  \label{tau_estim}

One can easily see that for an equation $\frac{\d f}{\d t} = \frac{\d}{\d x}
(A \frac{\d f}{\d x})$ the characteristic time of ``equillibrium establishment''
at spatial scale $l$ equals $\tau \sim \frac{l^2}{8A}$. 
Let us make a few estimates for kinetic equation (\ref{kin_eq})
with coefficients (\ref{R22b}, \ref{R11b}) calculated for the bulge.

The diffusion timescale for momenta $\tau_2 = \frac{(l_0 J_0)^2}{8R_{22}} \sim 10^6
\left(\frac{r_+}{1\mbox{ pc}}\right)^2$~yr, so that the galactic age corresponds 
to spatial area of $r_+ \le 100$~pc. 
The coefficient $R_{11}$ increases with decreasing $m$. The diffustion timescale
along $J$ axis at minimal momentum $m=m_g$ equals
$\tau_1 \sim 3\cdot 10^6 \frac{r_+}{1\mbox{ pc}}$~yr,
and for $r_+\sim 100$~pc $\tau_1$ is much less that the galactic age.
But for momenta at least an order of magnitude greater than $m_g$ the diffusion 
along $J$ axis does not disturb much the one-dimensional diffusion along $m$.
We will show in the next section that if we account for a cutoff of star
distribution function at  $r \to 0$, then we obtain a finite limit for $R_{11}$
with decreasing $m$. Thus $\tau_1$ becomes comparable with $\tau_2$
or greater for all $r_+ \le$ 100~pc.
Hence in the first approximation we can neglect the diffusion along $J$ in
comparison with the diffusion along $m$, at least for not very small values of 
$m$.

In addition we determine the characteristic timescale for diffusion in the
central region.
Since $R_{22} \sim J^2 = {l_0}^2 m^2$ the timescale does not depent on initial 
value of momentum and equals $\tau_2 = 10^6\mbox{ yr}\times
\left(\frac{M_{bh}}{3\cdot 10^6 M_\odot}\right)^2$.

\subsection{Diffusion coefficients for an improved star distribution}  \label{star_distrib}

Indeed, in deriving relations (\ref{F0conf}, \ref{F0coul}) we have neglected
the fact that number density of stars in the vicinity of black hole may be
significantly less than that follows from general power-law profile due to
stars capture or tidal disruption by black hole. 
From (\ref{R11b_int}) we see that it is the stars in the pericenter of the
orbit that affect the coefficient $R_{11}$ most of all. Hence for particles with 
low momenta these are the innermost stars, which number we may have overestimated. 
Now let us try to evaluate corrections linked to this fact.

Firstly, consider the Coulomb potential region. Observations show \cite{genzel3} 
that the nearest star's orbital axis approximately equals 1000 AU $=3\cdot 10^{-4}$ pc.
But from (\ref{center}) we see that the number of stars inside a sphere of radius
1000 AU should be about 15.

For correction we adopt that stars distribution function vanishes for energies
$E<E_{cr}=-\frac{G\,M_{bh}}{r_{cr}}$. Thus the density profile (\ref{center}) 
transforms to 
\begin{equation}  \label{centermod}
n(r) = n_0 \left(\frac{r}{r_0}\right)^{-3/2} 
  \left[1 - \left(1-\frac{r}{r_{cr}}\right)^{3/2}\right] \;.
\end{equation}

Comparison with observations gives the value $r_{cr} \simeq 5\cdot 10^{-3}$ 
pc $\sim 10^4 r_g$.

Now let us calculate modified coefficient $R_{11}$. Following (\ref{R11c_int}) 
we obtain the expression (with integration from 0 to 1 in this case):
\begin{equation}
R_{11} = \frac{16}{3\sqrt{2}} G^2 {M_s}^2 L_c F_0 r_+ |E|^{3/2} T 
\left\{ \begin{array}{ll}
(\frac{3\pi}{4} - \frac{\pi}{4}\chi_{cr})/\sqrt{\chi_{cr}} &, 
  \chi_{min}<\chi_{cr}<1 \vspace{10pt}\\
{\displaystyle \frac{\pi}{2\chi_{cr}}} &, \chi_{cr}>1 \end{array} \right.
\end{equation}
$$
\chi_{cr}=\frac{r_{cr}}{r_+}
$$
For $r_+<1$ pc, $\mu=\frac{m}{I} \le 0.1$, we have $\chi_{cr}>\chi_{min}$, 
hence in the whole central region of Coulomb potential the lower limit of 
integration $R_{11}$ is given by $\chi_{cr}$. This means that $R_{11}$ is in 
fact is much less (approximately $\frac{1}{3} \sqrt{\chi_{cr}/\chi_{min}}$ times) 
than calculated from (\ref{R11c}).

Secondly, the same formalism in the region of self-consistent (particularly, isothermal)
potential should include the statement that distribution function of stars vanishes
for sufficiently low energies. For isothermal star distribution this corresponds
to well-known solution for isothermal sphere with core (i.e. no central cusp), and
the core radius is of the same order that $r_{cr}$. Unfortunately, we can make
no estimation for core radius from observations, since their spatial resolution
is not enough. We may adopt the same value as calculated for our Galaxy 
($r_{cr} \sim 10^2 \div 10^4\, r_g$). One can show that this assumption changes 
the lower limit of integration in (\ref{R11b_int}) to value $\chi_{min} \sim 
max (r_-, r_{cr})/r_+$. Thus $R_{11}$ increases with decreasing momentum
up to $m=(10^2\div 10^3) m_g$ and then reaches constant limit.
At the same time $R_{22}$ is not affected by this cut since the integrand in
(\ref{R22b_int}) is small at small $\chi$. 

\section{Dark matter absorption and growth of the black hole}

\subsection{One-dimensional diffusion approximation}

It was shown in previous sections that coefficient $R_{22}$ does not depend
on $m$ in the bulge outside the central region of BH influence; coefficient
$R_{11} \sim R_{22} \left(\frac{J}{m}\right)^{\epsilon}$, 
$\epsilon \le 1$, and $R_{12} \sim R_{22} \frac{m}{J}$. 
Since at the initial moment $m=l_0J \ll J$ and our scope of interest lies 
in the domain of low momenta, we can take $m\ll J$ and disregard the term with
$R_{12}$. Furthermore, from the same arguments it follows that
$\frac{\d f}{\d J} \sim l_0 \frac{\d f}{\d m}$ at the initial moment, 
and since $R_{11}\sim R_{22}$ at $m=l_0J$, then $R_{11}\d f/\d J \ll R_{22}\,
\d f/\d m$ at not very low values of $m$. 
Finally, $f(m,J,t)=0$ at $m=m_g$, hence $\left.\frac{\d f}{\d J}\right|_{m=m_g}=0$, 
and $R_{11}\frac{\d f}{\d J}$ is limited at $m\to 0$. 
Also notice that in fact $R_{11}$ itself is limited at $m\to 0$, as noted previously.

To summarize, in the first approximation we leave in (\ref{Stf}) only the first term
and rewrite the kinetic equation (\ref{kin_eq}) as follows:

\begin{equation}  \label{kin_eq_m}
\frac {\d f}{\d t} = \frac{1}{m} R(J)\frac{\d}{\d m} \left( m\,
  \frac{\d f}{\d m}\right) \;,
\end{equation}
\begin{equation}  \label{boundary}
\mbox{ with boundary conditions }\quad
f|_{m=m_g}=0\;,\quad  m\frac{\d}{\d m}f |_{m=\infty} = 0 \;,\qquad
\end{equation}
initial conditions (\ref{f_0}) and diffusion coefficient
\begin{equation}  \label{Rtot}
R(J) = 0.46\, G\,M_s\,L_c\,{\sigma_0}^\frac{1}{1+\alpha/2}\,
  J^\frac{\alpha}{2+\alpha} \cdot \left\{ \begin{array}{lcl}
  1 &,& J>J_0(M_{bh}) \\
  \left(\frac{J}{J_0}\right)^2 &,& J<J_0  \;\;.\end{array}\right.
\end{equation}
\begin{equation}  \label{J0}
\mbox{Here }J_0 = \sqrt{G\,M_{bh}\,R_0/2} \propto {M_{bh}}^{\frac{1+\alpha/2}{1+\alpha}}
\end{equation}
is the boundary value separating the BH domain of influence and bulge itself
(as follows from (\ref{R_0}, \ref{Jcoul})).

The flux of dark matter through the surface $m=m_g$ is given by the expression
$$
S(t) = -(2\pi)^3 \int\hspace{-8pt}\int\hspace{-8pt}\int dJ\, dm \, dm_z\,
  \frac{\d f}{\d t} =
  -(2\pi)^3 \int\hspace{-8pt}\int dJ\, dm \, 2m
  \frac{1}{m}R\frac{\d}{\d m} m \frac{\d f}{\d m} =
$$
\begin{equation}  \label{S_total}
= (2\pi)^3 \int dJ\cdot 2 \left. \left(m R(J)\frac{\d f}{\d m}\right)\right|_{m=m_g}
= 2(2\pi)^3 \int dJ\, f_0J^{1/8}\, S_J(t) \;,
\end{equation}
where $S_J(t) = m_g R \frac{\d f}{\d m}$ is the flux through $m=m_g$ 
for equation (\ref{kin_eq_m}) with initial condition 
\begin{equation}  \label{f_init_1d}
f(m,t=0)=\delta(m^2 - {m_0}^2) \;,\qquad m_0 = l_0\, J
\end{equation}

Now we are going to calculate the time-dependent flux $S_J(t)$ of DM particles 
through the absorption boundary for each value of $J$ separately, i.e. one-dimensional 
flux, and then integrate it over $J$ assuming that the diffusion along $J$ is small.

\subsection{Flux in one-dimensional diffusion}  \label{dif1d}

Consider an auxilliary task: equation (\ref{kin_eq_m}) with boundary conditions
(\ref{boundary}) and initial condition (\ref{f_init_1d}) and determine the flux
$S_J(t) = m_g\,R\frac{\d f}{\d m}$.

The solution of eq.(\ref{kin_eq_m}) may be represented as
\begin{equation}  \label{sol}
f(m,t) = \int_0^\infty dm'\,G(m,m',t)\,f(m',0) \;,
\end{equation}
$$
G = \int_0^\infty d\lambda\,m'\,\exp(-\lambda\, R\, t)\,
  Z_\lambda(m, m_g)\,Z_\lambda(m',m_g)  \;\mbox{ is the Green function}\;,
$$ \newcommand{\sql}{\sqrt{\lambda}\,}
$$
Z_\lambda(m,m_g) = \frac{ J_0(\sql m_g)\, Y_0(\sql m) - J_0(\sql m)\, Y_0(\sql m_g)}
  {({J_0}^2(\sql m_g) + {Y_0}^2(\sql m_g))^{1/2}}
$$
is the orthogonal system of fundamental functions of the boundary problem 
(\ref{boundary}), $J_0$, $Y_0$ are Bessel functions of first and second kind of 0th order.

The initial conditions (\ref{f_init_1d}) give $\;f(m,t) = \frac{1}{2m_0}G(m,m_0,t)$.

One can easily show that
$$
\frac{\d}{\d m} Z_\lambda(m,m_g) = \frac{2}{\pi\,m_g} 
\frac{1}{({J_0}^2(\sql m_g) + {Y_0}^2(\sql m_g))^{1/2}} \;.
$$

Then the flux
$$
S_J(t) = m_g\,R \frac{1}{2m_0} \int_0^\infty d\lambda\, m_0\, 
\exp(-\lambda\, R\, t)\,Z_\lambda(m_0,m_g)\, 
\left.\frac{\d Z_\lambda(m,m_g)}{\d m}\right|_{m=m_g} =
$$ \newcommand{\sqe}{\sqrt{\eta}\,}
$$
= \frac{R\, H(x,y)}{\pi\,R\,t}\;,\;\;
H(x,y) = \int_0^\infty d\eta\,\exp(-\eta)\, \frac{J_0(\sqe x)\, Y_0(\sqe y) - 
  J_0(\sqe y)\, Y_0(\sqe x)}{{J_0}^2(\sqe x) + {Y_0}^2(\sqe x)}.
$$
Here we have changed the variables: $\eta=\lambda\,R\,t$, $x=m_g/\sqrt{R\,t}$, 
$y=m_0/\sqrt{R\,t}$.

One can show that $H(x,y) \simeq Z(x,y) \exp(-\zeta y^2)$ at $y \ge x+4$, 
$Z(x,y)$ weakly depends on its arguments, $\zeta\sim 5$.
Hence $S_J(t) \propto \frac{1}{t}\exp(-\frac{{m_0}^2}{\zeta R\,t})$.
To obtain exact form of the dependence, we take the following consideration.

\begin{figure}
$$\includegraphics{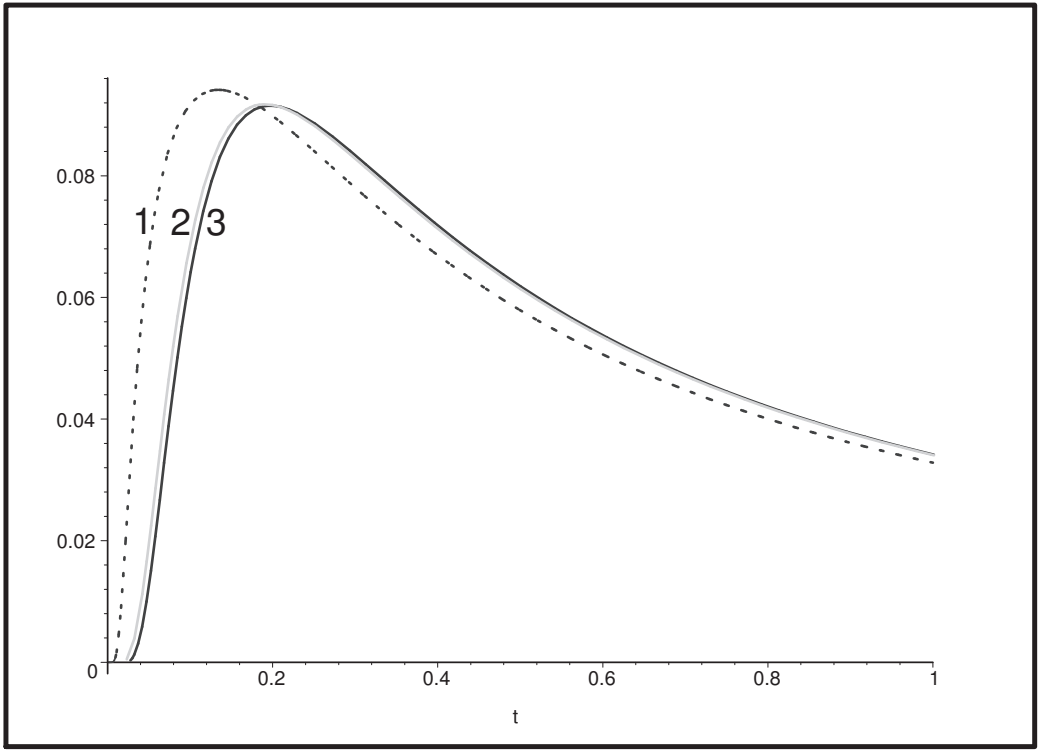} $$
\caption{ 
Normalized values $f(m,t)/\ln(m/m_g)$ for different $m$ (\ref{f_aux}): 
1) $m=m_0/2$, 2) $m=20m_g$, 3) $m=1.2m_g$. 
One can see that the graphs are quite similar, the normalized maximal 
values are approximately equal, though they are reached at different $t$.
\protect\\The values are given for $m_0=100m_g$.}  \label{f_scaled}
\end{figure}

The flux $S_J(t,m) = m\,R\,\frac{\d f}{\d m}$ is a continuous function of $m$; 
in the region $m_g<m<m_0$ we may take it to be a constant independent of $m$ 
(it is correct for values of $t$ greater than certain $t_0$ when the width of
peak of DF becomes comparable with $m_0$, see fig.\ref{f_scaled}). Let us denote
$ \displaystyle \kappa(t) = \left.\frac{\d f(m,t)}{\d m}\right|_{m={m_g}} $, 
then we have
\begin{equation}  \label{f_aux}
S_J(t)   = m_g\,R\,\kappa(t) \;,\;\; 
f(m,t) = \int_{m_g}^m \frac{S_J}{R\,m'} \d m' = m_g\,\kappa(t) \ln \frac{m}{m_g}\;.
\end{equation}

\begin{figure}
$$\includegraphics{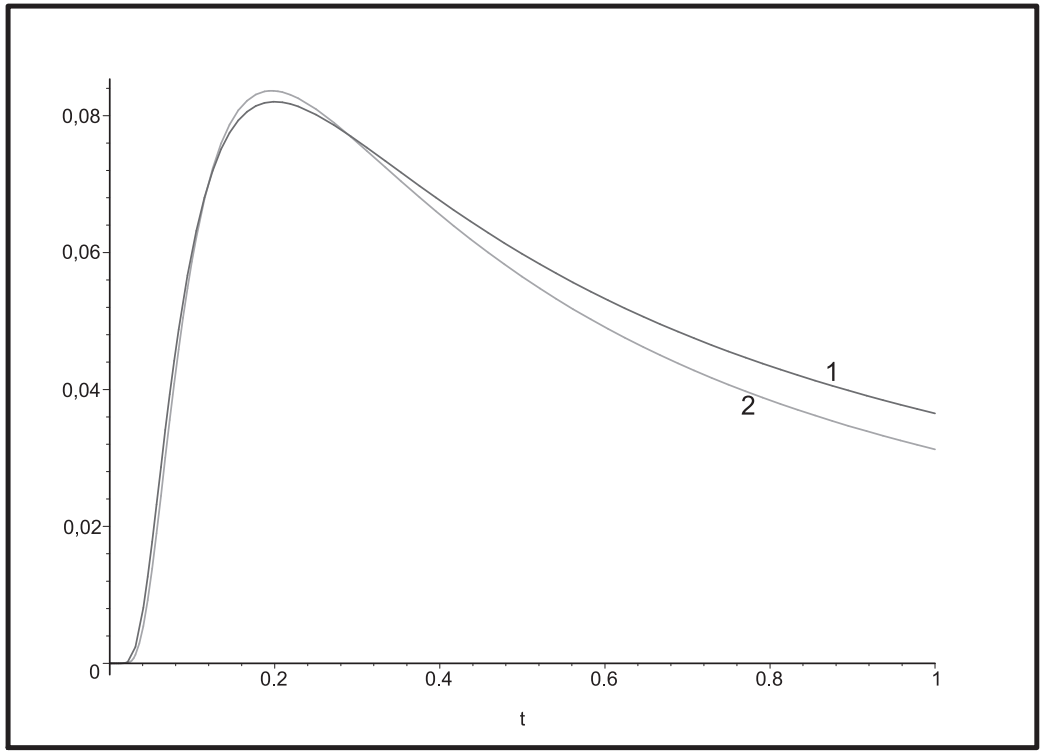} $$
\caption{ 
The flux function $\kappa(t)$: 1) theoretical (\ref{S_J}),
2) obtained by numerical integration. One can see that the theoretical 
approximation is good enough for maximum of the flux, but overestimates it at 
large $t$.
}  \label{kappa}
\end{figure}

Thus we obtain that $f(m,t)$ grows logarithmically with $m$. Its dependence on 
time can be found from the argument that $\kappa(t) \approx \frac{\Xi}{t}
\exp \left(-\frac{{m_0}^2}{5\,R\,t}\right)$, $\Xi$ is a constant.(fig.~\ref{kappa}). 
$\kappa(t)$ reaches maximal value at the exponent argument equal to $-1$; 
its maximal value $K_{max}=\frac{\Xi\,5R}{{\bf e}\, {m_0}^2}$. 
To determine the value of $\Xi$ we apply expression (\ref{f_aux}) for $m=m_0/2$;
having found analytically $f_{max}(m_0/2)=\frac{\ln 2}{2 {m_0}^2}$,
we obtain  
$\displaystyle \Xi=\frac{ {\bf e}\, \ln 2}{10\, m_g\,R\,\ln\frac{m_0}{2m_g}}$. 
Finally, the flux is
\begin{equation}  \label{S_J}
S_J(t) = \frac{ {\bf e}\, \ln 2}{10\, \ln\frac{m_0}{2m_g}} \cdot \frac{1}{t} 
\exp\left( - \frac{{m_0}^2}{5\,R(J)\,t}\right)
\end{equation}

\begin{figure}
$$ \includegraphics{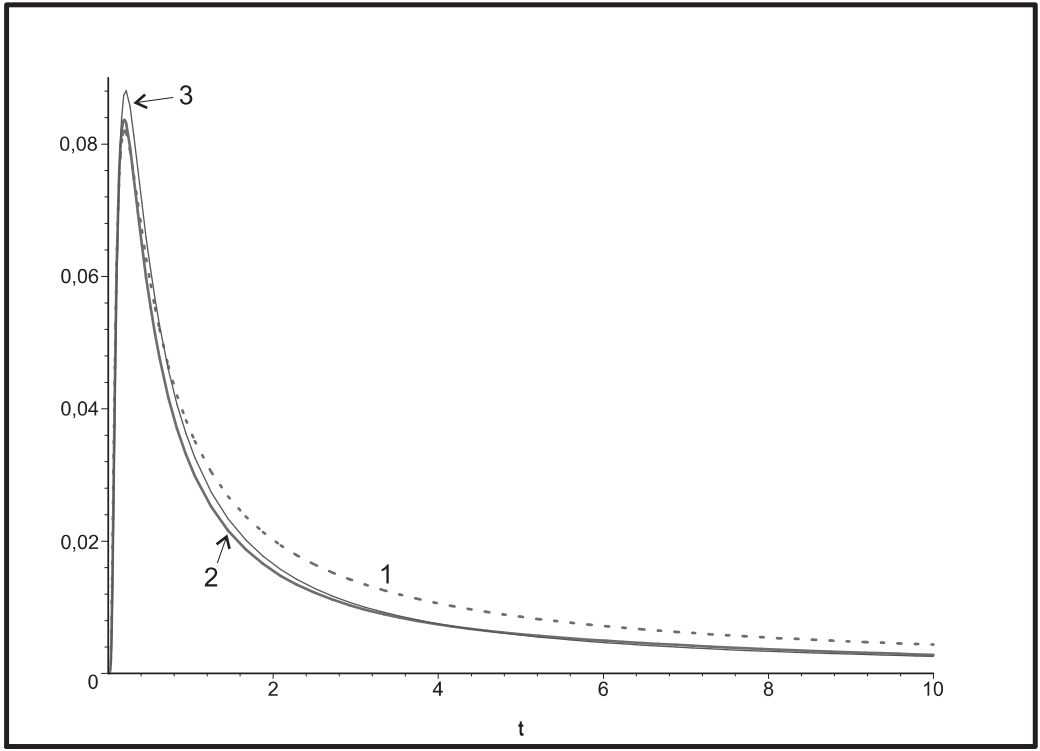} $$
\caption{ 
The corrected flux function $\kappa(t)$ for large $t$:
1) theoretical function (fig. \ref{kappa}),
2) obtained by numerical integration,
3) correcter theoretical function: $\kappa \propto t^{-5/4}$.
The last approximation is much better.}  \label{f_corrected}
\end{figure}

The correctness of the above consideration is proved by numerical investigation
of the problem. We also should notice that, in fact, expr$.$ (\ref{S_J})
slightly overestimates the flux at $t \ge \frac{{m_0}^2}{R(J)}$ (up to
a factor of 1.5, see fig.~\ref{f_corrected}).

\subsection{Black hole growth law}

Firstly, we neglect that in central region of Coulomb potential the coefficient
$R(J)$ differs from the expression (\ref{R22b})  and take $R(J) = 
R_\epsilon J^\epsilon$, $\epsilon=\frac{\alpha}{2+\alpha} \le 0.2$.
Substituting the obtained value (\ref{S_J}) of $S_J(t)$ into the expression 
(\ref{S_total}) for total flux, we rewrite it as follows:
\begin{equation}  \label{S_total2}
S(t) =
  \int_0^\infty f_0 J^{1/8} \frac{\scriptstyle 0.18}{\ln \frac{l_0J}{2m_g}} \frac{1}{t}
  \exp\left( -\frac{{l_0}^2 J^{2-\epsilon}}{5R_\epsilon t}\right) \;.
\end{equation}
The absorption boundary $m_g$ changes with time. However, the flux weakly 
depends on the value $m_g$. We approximate the logarithm in the denominator 
to have a constant value $\sim 10$.
\begin{equation}  \label{S_final}
S(t) = 8.9\,f_0\,H_\epsilon \frac{{R_\epsilon}^{\frac{9}{8(2-\epsilon)}}}
  {t^{1-\frac{9}{8(2-\epsilon)}}} \;,\quad
H_\epsilon = \left(\frac{5}{{l_0}^2}\right)^{\frac{9}{8(2-\epsilon)}} 
  \frac{\Gamma(\frac{9}{8(2-\epsilon)})}{2-\epsilon} \;.
\end{equation}

Supposing that black hole growth is governed only by the absorption of dark matter,
we obtain the following growth law for BH mass (assuming the mass of the seed
black hole to be small):
\begin{equation}  \label{Mbh_ot_t}
M_{bh}(t) = B\,(R_\epsilon\,t)^{\frac{9}{8(2-\epsilon)}} \;,\quad
B = 8.9\,f_0\,H_\epsilon\textstyle \frac{8(2-\epsilon)}{9}
\end{equation}

Notice that more precise expression for $S_J(t)$ at large $t$ reduces 
the value of $B$ approximately 1.2 times.

Thus the black hole growth is power-law with power index 
about $9/16$. This is in good agreement with previous work \cite{anton}, though
the power index is a bit lower.

\subsection{Influence of central region onto the black hole growth}

One could suppose that the diffusion goes slower in the central region since
the diffusion coefficient is lower there, and it may affect total growth law. 
In fact, nevertheless, it is not true. As has been shown in \S\ref{tau_estim},
the characteristic timescale for diffusion $\tau \propto {M_{bh}}^2$ and is about 
$10^6$ yr. at the moment, hence it was even lower previously. Additionally, from
(\ref{S_J}) one can see that maximum of the flux goes from inside the region
 $J_{max} \le \sqrt{5Rt/{l_0}^2}$, which at the moment corresponds to spatial 
area $r\le 100$~pc, and value of $J_0$ separating the central and outer parts of 
the bulge corresponds to $r\simeq 1$~pc. These two quantities depend on time in 
similar ways: $J_{max}\sim t^{1/2}$, $J_0\sim M_{bh}(t) \sim t^{9/16}$, 
so the relation $J_0 \ll J_{max}$ was also true in the past.

One can show that if $R$ varies in time, then we should take the following
expression for $S_J(t)$:
$$
S_J(t) = \frac{ {\bf e}\, \ln 2}{10\, \ln\frac{m_0}{2m_g}} \cdot \frac{R(t)}
{\int_0^t R(t')\,dt'} \exp\left( - \frac{l_0^2 J^2}{5\int_0^t R(t')\,dt'}\right)
\eqno (\ref{S_J}')
$$
For each $J$ we define $t_{max}(J)$ to be the time of maximal flux, and 
$t_0(J)$ to be the time to enter the BH domain of influence: $J_0(t_0)=J$.
The diffusion coefficient $R(t)$ starts to decrease after the time $t_0$. 
But one can easily see that 
$t_0(J)\simeq 10^4 \frac{{l_0}^2 J^2}{(5R)} = 10^4 t_{max}(J)$, 
from which we conclude that this correction affects only the far ``tail'' 
of $S_J(t)$, when then most part of the flux is already absorbed, 
and is practically unimportant for growth law.

\subsection{Influence of two-dimensional diffusion  onto the black hole growth}

Now let's try to estimate the effect of diffusion along $J$ axis, i.e. the 
correctness of one-dimensional approximation.
We restrict our consideration to the case $\alpha=0$ when $R_{22}=\rm const$.
Having obtained approximate solution for one-dimensional diffusion equation. we 
substitute it into initial equation (\ref{kin_eq}) and calculate the first term
$$
\eta = \frac{\d}{\d J}\left(R_{11} \frac{\d f}{\d J}\right) =
\frac{\d}{\d J}\left(R_{22}\,0.1\frac{J}{m}\, \frac{\d}{\d J} \left\{
  \frac{ {\bf e}\, \ln 2}{10\, \ln\frac{l_0J}{2m_g}} \frac{\ln\frac{m}{m_g}}{R_{22}t} 
\exp\left[ - \frac{{l_0}^2J^2}{5\,R_{22}\,t}\right] \right\}\right).
$$
Of special interest is the time $t\sim t_{max}$ corresponding to the maximum of 
the flux:
$t_{max} = \frac{{l_0}^2 J^2}{5R}$. It appears that $\eta \propto t_{max}-t$:
\begin{equation}  \label{eta_J}
\eta = \frac{\ln 2}{5\ln\frac{l_0J}{2m_g}}\, \frac{R_{22}\ln\frac{m}{m_g}}{{l_0}^2J^3m}
\,\left(1-\frac{t}{t_{max}}\right) \;.
\end{equation}
Now we should compare $\eta$ with the second term in (\ref{kin_eq}) 
$\eta_m=\frac{1}{m}\frac{\d}{\d m} \left(mR_{22}\frac{\d f}{\d m}\right)$. 
However, one can easily see that substituting (\ref{f_aux}) into this term makes it zero,
since this approximate solution does not satisfy initial equation in the whole region $m>m_g$. 
To avoid this problem, we calculate the left-hand side of (\ref{kin_eq_m})
$\eta_m=\frac{\d f}{\d t}$:
\begin{equation}
\eta_m = \frac{5\ln 2}{2\ln\frac{l_0J}{2m_g}}\, \frac{R_{22}\ln\frac{m}{m_g}}{{l_0}^4J^4}
\,\left(1-\frac{t}{t_{max}}\right)\;.
\end{equation}
Comparing with (\ref{eta_J}) we obtain $ \displaystyle
\frac{\eta}{\eta_m} = \frac{2\,{l_0}^2\,J}{25\,m}$.

For $J(r_+=100\mbox{ pc})$, $m=m_g$ this ratio is about 30. However, it was shown 
at the end of \S\ref{star_distrib} that to estimate maximal value of
$R_{11}$ one should take $m \sim 10^2 m_g$ instead of $m=m_g$, which lowers 
this ratio to about unity.  

It should be emphasized that all these estimations are rather approximate, 
since for precise assertions one should know the exact solution of 
one-dimensional diffusion. The arguments of \S\ref{dif1d} give only the 
correct value for flux through $m=m_g$, but not the exact solution for all $m$.

In general case, the inapplicability of reduction to one-dimensional diffusion at 
$m \to m_g$ does not change much the value of flux through $m=m_g$, since the 
diffusion along $J$ axis leads to the ``blur'' of distribution function along this 
axis, while the coefficient $R_{22}$ only weakly depends on $J$ as follows from
(\ref{R11b}). 
This is a benefit of selecting radial action as the second variable instead of 
energy as it was in previous works \cite{LS}.
However, an especial consideration is necessary whether
this diffusion may lead to the drift of dark matter into the
region of large $J$ where changes of $m,J$ during one period exceed their
values, and diffusion approximation becomes incorrect.

\section{Comparison with observations and conclusions}

In conclusion, we make theoretical estimations of black hole masses obtained from
growth law (\ref{Mbh_ot_t}) and compare them with observational data for 
several galaxies.

As it was noted previously, in distant galaxies it is difficult to measure precisely 
the dependence of velocity dispersion on radius. However, in galaxies M~31
and NGC~4258 it seems to be almost constant and equals approximately 200 km/s 
\cite{rotcurves}. Taking the value of $f_0$ the same as in our Galaxy (\ref{f_0})
and the time of growth $t=3\cdot10^{17}$~s $= 10^{10}$~yr. we obtain the value
$M_{bh} = 1.8\cdot 10^7 M_\odot$. The observed black hole masses are  
$(2.0 \div 8.5)\cdot 10^7 M_\odot$ for M~31 and $3.8\cdot 10^7 M_\odot$ for
NGC~4258 \cite{Tremaine}. The comparison shows that dark matter may comprise
significant fraction of black hole masses in these galaxies.

As for Milky Way, the rotation curve is not flat in the center of the bulge, and the 
velocity dispersion may be approximately represented as $\sigma(r) = \sigma_0 
\left(\frac{r}{10 \mbox{ pc}}\right)^{1/4}$, $\sigma_0=60$~km/s \cite{Tremaine}.
This corresponds to the value $\alpha=0.5$.
Using the assessment (\ref{f_0}) for $f_0$ the expression (\ref{Mbh_ot_t})
gives the black hole mass $M_{bh} \approx 10^7 M_\odot$, which clearly
overestimates the adopted value of $M_{bh} \approx 2.9\cdot 10^6 M_\odot$ 
\cite{genzel3} (about a factor of three). 
Notice that these values are about twice as smaller as in Ref.~\refcite{anton}, 
because of more precise estimate of dark matter flow (\ref{S_final}).
If we take $\sigma = \rm const = 100$~km/s, the numbers change only a little.

The disagreement with observations may be explained by rather rough estimates 
of quantities $f_0$ and $l_0$, which are determined by the whole dark matter 
halo and are different for different galaxies. 
If we take the value for DM fraction in bulge $\Upsilon \sim 0.24 \div 0.31$
\cite{Klypin2}, then the resulting BH mass will be almost in agreement
with observation.

In conclusion, we can say that the model discussed can give reasonable estimate 
for observed masses of giant black holes in galactic centers. It is likely that 
a large fraction of black hole mass may be comprised of dark matter.

Further development of this problem will require, firstly, more precise 
calculation of diffusion coefficients based on detailed data for star 
distribution in the central parts of the bulge; secondly, taking into account 
the bulge evolution; and finally, an exact consideration of two-dimensional 
diffusion accounting for particles Fermi-heating and their drift out of bulge.

The authors are grateful to A.V.~Gurevich, K.P.~Zybin, A.S.~Ilyin and V.A.~Sirota
for numerous fruitful discussions. This work was supported by RFBR grants 
01-02-17829, 03-02-06745, Russian Ministery of Education and Science grant 
2063.2003.2, and the Forschungszentrum Juelich grant 
within the framework of the Landau Program.


\newpage

\newpage

\newpage

\newpage

\end{document}